\documentclass[a4paper,12pt,english]{article}
\usepackage{amsfonts,bm,amssymb,euscript,array,babel}
\usepackage{amsthm}
\usepackage{mathtools}
\usepackage{tikz-cd}
\usepackage{amsmath}
\usepackage{hhline}
\usepackage[amsmath]{empheq}
\usepackage{hyperref}
\usepackage{cancel}
\usepackage{empheq}
\usepackage{fancybox}
\usepackage{mathrsfs}
\usepackage{pdfsync}
\usepackage{tikz-cd}

\newsavebox{\mysaveboxM}
\newsavebox{\mysaveboxT}

\makeatletter

\newcommand{\w}{\wedge}
\newcommand{\bbm}{\left(\begin{matrix}}
	\newcommand{\ebm}{\end{matrix}\right)}
\newcommand{\beq}{\begin{eqnarray}}
	\newcommand{\eeq}{\end{eqnarray}}

\makeatother

\newtheorem{theorem}[subsection]{Theorem}

\newtheorem{defn}[subsection]{Definition}
\newtheorem{rmk}[subsection]{Remark}

\newcommand{\sfrac}[2]{{\textstyle\frac{#1}{#2}}}

\newcommand{\be}{\begin{equation}}
	\newcommand{\ee}{\end{equation}}

\newcommand{\beqa}{\begin{eqnarray}}
	\newcommand{\eeqa}{\end{eqnarray}} 
\def\nn{\nonumber} \def \bea{\begin{eqnarray}} \def\eea{\end{eqnarray}}

\newcommand{\barr}{\begin{array}}
	\newcommand{\earr}{\end{array}}
\numberwithin{equation}{section}

\def\a{\alpha}  \def\b{\beta}

\def\s{\sigma}  \def\t{\tau} 

 \def\one{\mbox{1 \kern-.59em {\rm l}}}

\def\bit{\begin{itemize}} \def\eit{\end{itemize}}

\def\({\left(} \def\){\right)}

 \allowdisplaybreaks[3]

\begin{document}
\renewcommand{\title}[1]{\vspace{10mm}\noindent{\Large{\bf
			
			#1}}\vspace{8mm}} \newcommand{\authors}[1]{\noindent{\large
		
		#1}\vspace{5mm}} \newcommand{\address}[1]{{\itshape #1\vspace{2mm}}}


\begin{titlepage}
	
	
	\begin{center}
		
		
		\title{ {\Large From Hopf algebra to braided $L_\infty$-algebra }}
		
		\vskip 3mm
		
		  \authors{ 
		\large
		    Clay J. Grewcoe{\footnote{cgrewcoe@yahoo.com.au}} , \ 
		    Larisa Jonke{\footnote{larisa@irb.hr}} , \\ \vspace{10pt} 
		   Toni Kod\v zoman{\footnote{toni.kodzoman@irb.hr}} , \ 
		 George Manolakos{\footnote{giorgismanolakos@gmail.com}}}
		 
		 \vskip 3mm
		 
		  \address{ Division of Theoretical Physics, Rudjer Bo\v skovi\'c Institute \\ Bijeni\v cka 54, 10000 Zagreb, Croatia }
		
		
		\begin{abstract}
			\noindent
			We show that an $L_\infty$-algebra can be extended to a graded   Hopf algebra with a  codifferential. Then we  twist this extended  $L_\infty$-algebra with a Drinfel'd twist,  simultaneously twisting its modules.  Taking the $L_\infty$-algebra  as its own (Hopf)  module,  we obtain the recently proposed braided $L_\infty$-algebra. 
			The Hopf algebra morphisms are identified with the strict  $L_\infty$-morphisms, while  the braided $L_\infty$-morphisms define a more general $L_\infty$-action of twisted $L_\infty$-algebras.
			
		\end{abstract}
		
	\end{center}
	
	\vskip 2cm
	
\end{titlepage}

%

\section{Introduction}

$L_\infty$-algebras or homotopy Lie algebras are generalizations of Lie algebras with infinitely-many higher brackets, related to each other by higher homotopy versions of the Jacobi identity \cite{js1,csft,ls}. From a physical point of view, these higher algebras  represent the geometrical structure providing deeper understanding of  quantization of  field theory and gravity. In particular, there exists a  correspondence between the BV-formalism used in the quantization of gauge theories and $L_\infty$-algebras, as noticed by Zwiebach in his seminal work on closed string field theory \cite{csft}.   Recently, the BV  $L_\infty$-algebra of Yang-Mills  theory was used in relating the square of gluon amplitudes to that  of (${\cal N}=0$ super) gravity amplitudes to all orders of perturbative quantum field theory \cite{jurcodc}. On the other hand, in the framework of  deformation quantization   \cite{bffls}, Kontsevich's  famous  formality theorem on the existence and classification of star products on Poisson manifolds has been proven using the concept of  $L_\infty$-quasi-isomorphisms  \cite{mk}. Subsequently, Cattaneo and Felder \cite{cf} provided an interpretation of  the Kontsevich quantization  formula in terms of the perturbative expansion of the path integral for the Poisson sigma model \cite{ikeda, thomas}.

 More recently,  attempts to improve the understanding of  consistent non-commutative deformations of field theory  using an $L_\infty$-framework resulted in two interesting proposals. In Ref.\cite{Vlad}, the homotopy relations defining an  $L_\infty$-algebra were used to bootstrap a consistent star-gauge invariant theory starting from the star-deformed commutator of the symmetry algebra.     However, there is another consistent way of introducing non-commutative deformations based on Drinfel'd twists of the symmetry Hopf algebra  \cite{dt}. The analysis of twisted gauge symmetries was initiated in Refs.\cite{t0,t1,t2,t3}, but only recently  was it shown that  these twisted   symmetries can be understood  in the framework of  braided $L_\infty$-algebras \cite{Marija,GR}. The construction of braided $L_\infty$-algebras  in \cite{Marija} was based on  a reinterpretation of  all  the defining relations of an $L_\infty$-algebra in terms of morphisms in a suitable category. 
 By twisting the enveloping Hopf algebra of vector fields on a manifold M to a non-cocommutative  Hopf algebra and simultaneously deforming its category of modules,  the $L_\infty$-algebra was deformed into a braided $L_\infty$-algebra.  
 Here we show that an $L_\infty$-algebra itself can be extended to  a graded   Hopf algebra with a compatible codifferential, and using  this observation we construct the   braided $L_\infty$-algebra using a Drinfel'd twist of its  underlying Hopf algebra structure.

In the next section  we recall the coalgebra structure of an $L_\infty$-algebra   and show that it can be extended to a graded   Hopf algebra with a codifferential.   In section 3 we apply Drinfel'd twists to the Hopf algebras underlying $L_\infty$-algebras and obtain twisted $L^{\cal F}_\infty$-algebras.   In the spirit of deformation quantization, after twisting the algebra, one further twists its modules.  Taking the Hopf algebra  as its own module,  we obtain another Hopf algebra, $L^\star_\infty$, which is exactly the braided  $L_\infty$-algebra defined in \cite{Marija}. Furthermore,  we reinterpret the Hopf algebra (iso)morphism  in terms of strict $L_\infty$-(iso)morphisms, and define a more general class of  braided $L_\infty$-morphism. In the conclusion, we briefly discuss the relevance of our results for the braided gauge field theory.

\section{$L_\infty$  as  Hopf algebra}

An $L_\infty$-algebra can be defined in several different ways, depending on the context.  In the standard approach one defines an $L_\infty$-algebra on  a graded vector space $X$ as a generalization of a Lie algebra  with possibly infinitely-many higher brackets, related to each other by higher homotopy versions of the Jacobi identity \cite{js1,csft,ls}. Alternatively, one can describe an $L_\infty$-structure as a degree 1 coderivation  on the coalgebra generated by the suspension\footnote{The suspension map is also called a shift isomorphism \cite{Jurco}, see Appendix A for more details.} of $X$, as shown in Refs.\cite{ls, LadaM}. In order to identify the Hopf algebra structure underlying an $L_\infty$-algebra this  coalgebra picture is more appropriate and we shall review it here\footnote{In the rest of the  paper we work in the coalgebra picture and denote with $X$  the underlying graded vector space  to simplify notation.}.

 \noindent Let us start  with a graded symmetric tensor algebra:
$${\mathbf{S}}(X):=\bigoplus_{n=0}^\infty S^nX~,$$
where $X$ is a $\mathbb{Z}$-graded vector space $X=\bigoplus_{d\in \mathbb{Z}} X_d$ over the field $K=S^0X$ and the degree of a homogeneous element $x_i\in X$ is denoted as $|x_i|$. The tensor product $\mu:{\mathbf{S}}(X)\otimes {\mathbf{S}}(X)\to {\mathbf{S}}(X)$  is graded  symmetric, 
$$\mu(x_1\otimes x_2)=x_1\vee x_2=(-1)^{|x_1||x_2|}x_2\vee x_1=(-1)^{|x_1||x_2|}\mu(x_2\otimes x_1),\qquad x_1,x_2\in X~,$$
and we use $\vee$ to denote the product in ${\mathbf{S}}(X)$. The algebra structure can be endowed with a unit map $\eta:K\to{\mathbf{S}}(X)$, where $\eta(1)=1$.  

The coalgebra structure on ${\mathbf{S}}(X)$ is given by the coproduct:
\bea\label{nlab}
\Delta(x_1\vee\cdots\vee x_m)= \sum_{p=0}^{m}\sum_{\sigma\in \mathrm{Sh}(p, m-p)}\!\!\!\!\epsilon(\sigma;x)(x_{\sigma(1)}\vee\cdots\vee x_{\sigma(p)})\otimes (x_{\sigma(p+1)}\vee\cdots\vee x_{\sigma(m)})~,
\eea
where $\epsilon(\sigma;x)$ is the Koszul sign,
$$x_1\vee \cdots \vee x_k=\epsilon(\sigma;x)x_{\s(1)}\vee \cdots \vee x_{\s(k)},\qquad x_i\in X~,$$
and  $\mathrm{Sh}(p, m-p)\in S_m$ denotes those permutations ordered as $\sigma(1)<\cdots<\sigma(p)$ and $\sigma(p+1)<\cdots<\sigma(m)$. We use the conventions that Sh$(n,0)=\mathrm{Sh}(0,n)$ equals id $\in S_n$
and that an empty slot in the product equals the unit,  $1\in K$.  Thus we have
\begin{align}
\Delta(1)&=1\otimes 1~,\nn\\
\Delta(x)&=1\otimes x+x\otimes 1~,\nn\\
\Delta(x_1\vee x_2)&=1\otimes(x_1\vee x_2)+(-1)^{|x_1||x_2|}x_2\otimes x_1+x_1\otimes x_2+(x_1\vee x_2)\otimes 1~,\nn\\
&\cdots\nn\end{align}
As a map $\Delta: {\mathbf{S}}(X)\to {\mathbf{S}}(X)\otimes {\mathbf{S}}(X)$, this reads:
\bea\label{map}
&& \Delta\circ \mathrm{id}^{\vee m}= \sum_{p=0}^{m}\sum_{\sigma\in \mathrm{Sh}(p, m-p)}(\mathrm{id}^{\vee p}\otimes \mathrm{id}^{\vee(m-p)})\circ \t^\sigma~,\qquad p,m\geq 0~,
\eea
where the  $\t^\sigma$ denotes the action of permutations  \cite{Marija}, e.g. the non-identity permutation of two elements is:
$$\t^\s(x_1\vee  x_2)=(-1)^{|x_1||x_2|}x_2\vee x_1~,$$
and includes the Koszul sign.  Furthermore, the coalgebra structure on ${\mathbf{S}}(X)$ includes counit $\varepsilon:{\mathbf{S}}(X)\to K$, where $\varepsilon(1)=1$ and $\varepsilon(x)=0,\;x\in X$.

Next,   we introduce a coderivation $D$ that squares to zero and thus generates the appropriate homotopy relations. The coderivation is a map $D:{\mathbf{S}}(X)\to {\mathbf{S}}(X)$  of degree 1 such that the co-Leibniz property is satisfied,
\begin{equation}
    \Delta\circ D=(1\otimes D +D\otimes 1)\circ \Delta~.
\end{equation}
This  coderivation is given as \cite{ls}:
\be
D=\sum_{i=0}^{\infty}b_i~,
\ee
where the graded symmetric multilinear maps $b_i$ are of degree $1$. When $b_0$ is non-vanishing one talks about curved $L_\infty$-algebras \cite{curved1,curved2}, while for $b_0=0$ we have flat $L_\infty$-algebras\footnote{We shall use the term $L_\infty$-algebra for both cases when the distinction is not relevant.}. The $b_i$ maps act on the full tensor algebra as a coderivation:
\begin{align} \label{mapbi}
&b_i:S^jX\to S^{j-i+1}X~,\\
b_i(x_1\vee\ldots\vee x_j)=\sum_{\sigma\in \mathrm{Sh}(i, j-i)}\epsilon(\sigma;x&)b_i(x_{\sigma(1)},\ldots , x_{\sigma(i)})\vee x_{\sigma(i+1)}\vee\ldots\vee x_{\sigma(j)},\qquad j\geq i~,\nn
\end{align}
and can be written using the permutation map $\tau^{\s}$ as:
\bea \label{mapbimap}
b_i\circ\mathrm{id}^{\vee j}=\sum_{\sigma\in \mathrm{Sh}(i, j-i)} (b_i\vee \mathrm{id}^{\vee (j-i)}) \circ \t^\sigma~,\qquad j\geq i~.
\eea
\noindent Note that $b_0(1)=b_0$ is a degree 1 element of $S^1X=X$. Now one can define an $L_\infty$-algebras as a $\mathbb{Z}$-graded vector space with multilinear graded symmetric maps $b_i:X^{\otimes i}\to X$ of degree $1$ such that the coderivation $D=\sum_{i=0}^{\infty}b_i$ is nilpotent \cite{LadaM}.  As an example, we calculate  the first few homotopy relations:
\begin{align}
D^2&(x_1\vee x_2)=\sum_{i=0}^{\infty}b_i\sum_{j=0}^{2}b_j(x_1\vee x_2)\nn\\
&=\sum_{i=0}^{3}b_i(b_0\vee x_1\vee x_2+b_1(x_1)\vee x_2+(-1)^{|x_1||x_2|}b_1(x_2)\vee x_1+b_2(x_1,x_2))\nn\\
&=b_1(b_0)\vee x_1\vee x_2 +\nn\\
&\phantom{\,=\,}+b_1^2(x_1)\vee x_2+(-1)^{|x_1||x_2|}b_1^2(x_2)\vee x_1+b_2(b_0,x_1)\vee x_2+(-1)^{|x_1||x_2|}b_2(b_0,x_2)\vee x_1+\nn\\
&\phantom{\,=\,}+b_1(b_2(x_1,x_2))+b_2(b_1(x_1),x_2)+(-1)^{|x_1||x_2|}b_2(b_1(x_2),x_1)+b_3(b_0,x_1,x_2)~.\nn
\end{align}
The vanishing of the above expression is equivalent to the following three identities:
\begin{align}
 b_1b_0&=0~, \nn\\
b_2b_0+b_1^2&=0~, \label{bhom}\\
b_3b_0+b_2b_1+b_1b_2&=0~. \nn
\end{align}
Additionally, we used the fact that $b_0^2$ is trivially zero due to the odd degree of $b_0$, and is therefore not a constraint.
The homotopy relations defining an $L_\infty$-algebra can be written in the closed form \cite{Jurco}:
\begin{equation}\label{closed}
\sum_{j=0}^i \sum_{\sigma\in \mathrm{Sh}(j,i)} b_{i-j+1}(b_j\vee \mathrm{id}^{\vee i})\circ\tau^\sigma=0~. \end{equation}
 Moreover, we have $\varepsilon\circ D=0$ as $\varepsilon(b_0(1))=0$.

 So far we have identified an  $L_\infty$-structure with a  counital coalgebra over a graded vector space with compatible  coderivation that squares to zero.    Now we wish to compare this structure with the one of a Hopf algebra. In short, a Hopf algebra is a bialgebra that admits an antipode map with  certain compatibility properties. While the formal definition is given in Appendix B,  we discuss here the  prototypical example -- a tensor algebra. 

\noindent{A tensor algebra} $T(V)=\bigoplus_{n=0}^\infty T^nV$, where $V$ is a vector space over the field $K$  can be seen as a Hopf algebra $(T(V),\cdot,\Delta,\varepsilon,S)$. The coproduct $\Delta$, counit $\varepsilon$  and antipode $S$ are defined on $v\in V$ as:
\begin{align*}
\Delta(v)&=v\otimes 1+1\otimes v~,\mkern-100mu &\Delta(1)&=1\otimes 1~,\\
\varepsilon(v)&=0~, &\varepsilon(1)&=1~,\\
S(v)&=-v~,\mkern-100mu & S(1)&=1~. 
\end{align*}
Since the coproduct   and counit   are algebra homomorphisms and the antipode is an algebra (and coalgebra) anti-homomorphism, we can extend the definition from the basis elements to the full tensor algebra:
\begin{align}
S(v_1\cdot\ldots\cdot v_m)&=(-1)^m v_m\cdot\ldots\cdot v_1~,\nn\\
 \Delta(v_1\cdot\ldots\cdot v_m)&= \sum_{p=0}^{m}\sum_{\sigma\in \mathrm{Sh}(p, m-p)}(v_{\sigma(1)}\cdot\ldots\cdot v_{\sigma(p)})\otimes (v_{\sigma(p+1)}\cdot\ldots\cdot v_{\sigma(m)})~,\nn
\end{align}
where we use $\,\cdot\,$ for the product in the tensor algebra. This example can be trivially extended to the symmetric graded tensor algebra used in the construction of an $L_\infty$-algebra above.  In particular, the coproduct will be of the form \eqref{nlab}, and the antipode will be extended to the graded antipode:
\bea \label{gS}
S(x_1\vee\cdots\vee x_m)=(-1)^m(-1)^{\sum_{i=2}^m\sum_{j=1}^{i-1}|x_i||x_j|}x_m\vee\cdots\vee x_1.
\eea
Using the axioms of a Hopf algebra given in Appendix B, one can easily verify that the symmetric graded tensor algebra is indeed a Hopf algebra.
 Thus we arrive to the following theorem.
\begin{theorem} 
An extended $L_\infty$-algebra is a bialgebra $({\mathbf{S}}(X),\mu,\eta,\Delta,\varepsilon)$ with coderivation $D: {\mathbf{S}}(X)\to {\mathbf{S}}(X)$ of degree 1 s.t.
the  co-Leibniz property is satisfied
  $$\Delta\circ D=(1\otimes D +D\otimes 1)\circ \Delta~,$$ and $D^2=0$.
   It naturally inherits the structure of a Hopf algebra from the graded symmetric tensor algebra, with:
   \begin{equation*}
   	S\circ D=\widetilde{D}\circ S, \quad \varepsilon\circ D= 0, 
   \end{equation*}
   \noindent where the codifferential $\widetilde{D}$ 
   \begin{equation*}
   	\widetilde{D}=\sum_{i=0}^\infty\tilde{b}_{i}=\sum_{i=0}^\infty(-1)^{1-i}\,b_{i}~,
   \end{equation*}
induces the same homotopy relations as $D$.
\end{theorem}
\noindent Note that the unit map $\eta:K\to {\mathbf{S}}(X)$ is in general a morphism of graded coalgebras, and only for a flat  $L_\infty$-algebra, i.e., when $b_0=0$ does it become a morphism of differential graded coalgebras with $D\circ\eta=0$.  
\begin{proof}
We need to show that the Hopf algebra structure of the symmetric graded tensor algebra is compatible with the $L_\infty$-algebra structure encoded in the nilpotent coderivation. The compatibility of the unit and counit with coderivation was already discussed, therefore we need to check only the antipode compatibility relation. We apply the coderivations $b_i$ and $\tilde b_i$ on an element of $S^jX$ using \eqref{mapbi}. The left hand side is 
\bea
&&S(b_{i}(x_{1}\vee\ldots\vee x_{j})) = \sum_{\sigma\in \mathrm{Sh}(i, j-i)}\epsilon(\sigma;x)S(b_i(x_{\sigma(1)},\ldots , x_{\sigma(i)})\vee x_{\sigma(i+1)}\vee\ldots\vee x_{\sigma(j)})\nn\\
&&= \sum_{\sigma\in \mathrm{Sh}(i, j-i)}\epsilon(\sigma;x)(-1)^{\cal P}S(x_{\sigma(j)})\vee\ldots\vee S(x_{\sigma(i+1)})\vee S(b_i(x_{\sigma(1)},\ldots , x_{\sigma(i)}))\nn\\
&&= \sum_{\sigma\in \mathrm{Sh}(i, j-i)}\epsilon(\sigma;x)(-1)^{j-i+1}(-1)^{\cal P}x_{\sigma(j)}\vee\ldots\vee x_{\sigma(i+1)}\vee b_i(x_{\sigma(1)},\ldots , x_{\sigma(i)})\nn\\
&&= \sum_{\sigma\in \mathrm{Sh}(i, j-i)}\epsilon(\sigma;x)(-1)^{j-i+1}b_{i}(x_{\sigma(1)},\ldots , x_{\sigma(i)})\vee x_{\sigma(i+1)}\vee\ldots\vee x_{\sigma(j)}~,\nn
\eea
where in the second line we introduced the sign 
$${\cal P}=\(\sum_{m=1}^i|x_{\s(m)}|+1\) \sum_{n=i+1}^{j}|x_{\s(n)}|+\sum_{m=i+2}^j\sum_{n=i+1}^{m-1}|x_{\s(n)}||x_{\s(m)}|~,$$ 
induced by the graded action of antipode \eqref{gS}  and in the third line we used $S(x)=-x,\;\forall x\in X$. Similarly, the right hand side gives
\begin{align*}
&	\tilde b_i(S(x_{1}\vee\ldots\vee x_{j})) = \tilde b_{i}((-1)^{\cal\widetilde P}S(x_{j})\vee\ldots\vee S(x_{1}))= (-1)^{j}\tilde  b_{i}(x_{1}\vee\ldots\vee x_{j})\\
&= \sum_{\sigma\in \mathrm{Sh}(i,j-i)}\epsilon(\sigma;x)(-1)^{j}\tilde b_i(x_{\sigma(1)},\ldots , x_{\sigma(i)})\vee x_{\sigma(i+1)}\vee\ldots\vee x_{\sigma(j)}~,
\end{align*}
where ${\cal\widetilde P}=\sum_{m=2}^j\sum_{n=1}^{m-1}|x_n||x_m|$.
Equating the two sides gives $\tilde b_i=(-1)^{1-i}b_i$. Inspecting the homotopy relations \eqref{bhom} it is easy to see that the  relations induced by $\widetilde D^2=0$ are the same.
\end{proof}
\noindent A coderivation of  a graded Hopf algebra  with similar properties was previously introduced in   Ref.\cite{Schupp} in the context of the BRST formulation of quantum group gauge theory.

\section{Braided $L_\infty$-algebra from a Drinfel'd twist}

The Hopf algebra underlying   an $L_\infty$-algebra we have discussed so far is cocommutative and coassociative, as it is based on a (graded) symmetric tensor algebra. A systematic way to introduce a non-(co)commutative deformation is by applying  the Drinfel'd twist approach \cite{dt}. We twist a Hopf algebra $H$ using a twist element ${\cal F}\in H{\otimes}H$, which is invertible and satisfies:
\begin{align}
 ({\cal F}\otimes 1)(\Delta\otimes \mathrm{id}){\cal F}&=(1\otimes {\cal F})( \mathrm{id}\otimes \Delta){\cal F}~, \label{cc1}\\
(\varepsilon\otimes \mathrm{id}){\cal F}=1\otimes 1&=( \mathrm{id}\otimes\varepsilon){\cal F}~.\label{cc2}
\end{align}
 Relation (\ref{cc1}) is known as the 2-cocycle condition, whereas the condition (\ref{cc2}) is   known as (normalized) counitality. The 2-cocyle condition ensures that the deformed algebra remains coassociative. 
Using Sweedler's summation notation  we can   write the twist element and its inverse  as:
\begin{equation}
    \mathcal{F}=f^{\alpha}\otimes f_{\alpha},\qquad\mathcal{F}^{-1}=\bar{f}^{\alpha}\otimes\bar{f}_{\alpha}~.
\end{equation}
It was shown in Refs.\cite{Majid,admw} that a twist $\mathcal{F}$ of a Hopf algebra $H$ results in a new Hopf algebra $H^{\mathcal{F}}$ which is given by $(H,\mu,\Delta^{\mathcal{F}},\varepsilon, S^{\mathcal{F}})$. On the level of vector spaces $H^{\mathcal{F}}=H$, the product $\mu$ and counit $\varepsilon$ are unchanged, while the coproduct transforms as:
\begin{align}
\Delta^{\mathcal{F}}(h)=\mathcal{F}\Delta(h)\mathcal{F}^{-1},\qquad h\in H~.
\end{align}
 In the case of an Abelian twist\footnote{The twist generators  commute in the case of an Abelian twist.}, which we  assume in the following,  the antipode is not deformed, $S^{\cal F}=S$. Thus using the Drinfel'd  twist we obtain a twisted $L_\infty$-algebra, namely $(L_\infty^{\cal F}, \vee, \Delta^{\cal F},\varepsilon,S)$, where $L_\infty^{\cal F}$ and $L_\infty$ are the same as vector spaces.
 
 Drinfel'd twist deformation quantization consists of twisting the  Hopf algebra  as above, while simultaneously twisting all of its modules \cite{book}. Taking the Hopf algebra $L_\infty$ as a module  itself, one  obtains another Hopf  algebra
$(L_\infty^\star, \vee_\star, \Delta_\star,\epsilon,S_\star)$ with the corresponding vector space once again being the same as before, namely $L_{\infty}$, with the following product:
\bea
x_1\vee_\star x_2=\bar f^\alpha(x_1)\vee \bar f_\alpha(x_2)~.\eea
This algebra is a Hopf algebra with:
\begin{align}\label{starD}
\Delta_\star(x)&=x\otimes 1+\bar R^\alpha\otimes \bar R_\alpha(x)~,\\
 S_\star(x)&=-\bar R^\alpha(x) \bar R_\alpha~.\label{starS}
\end{align}
The $\cal R$-matrix ${\cal R}\in {\bold S}(X)\otimes {\bold S}(X)$ is  an invertible matrix induced by the twist,
\bea
{\cal R}={\cal F}_{21}{\cal F}^{-1}=:R^\alpha\otimes R_\alpha~,\; 
{\cal R}^{-1}=\bar R^\alpha\otimes \bar R_\alpha\eea
where $ {\cal F}_{21}=f_\alpha\otimes f^\alpha$.  In the case of an Abelian twist,  ${\cal R}$ is triangular $R_\alpha\otimes R^\alpha=\bar R^\alpha\otimes \bar R_\alpha$, and  ${\cal R}={\cal F}^{-2}$. The inverse $\cal R$-matrix controls the non-commutativity of the $\vee_\star$-product
and provides a representation of the permutation group \cite{book} and, in particular, 
the action of a non-identity permutation of two elements is:
$$\t^\s_R(x_1\vee_\star x_2)=(-1)^{|x_1||x_2|}\bar R^\alpha (x_2)\vee_\star \bar R_\alpha(x_1)~.$$
As ${\cal R}$ is triangular, $\t^\sigma_R$ squares to the identity. 
Now we can extend the coproduct \eqref{starD} to the whole tensor algebra:
\bea\label{mapstar}
&& \Delta_\star\circ \mathrm{id}^{\vee_\star m}= \sum_{\sigma\in \mathrm{Sh}(p, m-p)}(\mathrm{id}^{\vee_\star p}\otimes \mathrm{id}^{\vee_\star(m-p)})\circ \t^\sigma_R~,\qquad p,m\geq 0~.
\eea
The   coderivation $D_\star=\sum_{i=0}^\infty b_i^\star$
is defined in terms of braided graded symmetric maps $b_i^\star$:
\begin{align} \label{mapbimapstar}
b_i^\star\circ \mathrm{id}^{\vee_\star j}&=\sum_{\sigma\in \mathrm{Sh}(i, j-i)} (b_i^\star\vee_\star \mathrm{id}^{\vee_\star (j-i)}) \circ \t^\sigma_R~,\qquad j\geq i~,\\
b_i^\star(x_1,\ldots,x_m,x_{m+1},\ldots,x_i)&=(-1)^{|x_m||x_{m+1}|}b_i^\star(x_1,\ldots,\bar R^\a(x_{m+1}),\bar R_\a(x_{m}),\ldots,x_i)~,\nn
\end{align}
with the condition  $D_\star^2=0$ reproducing the deformed homotopy relations. In particular we have:
\begin{align}
D_\star^2&(x_1\vee_\star x_2)=\sum_{i=0}^{\infty}b^\star_i\sum_{j=0}^{2}b^\star_j(x_1\vee_\star x_2)\nn\\
&=\sum_{i=0}^{3}b^\star_i(b^\star_0\vee_\star x_1\vee_\star x_2+b^\star_1(x_1)\vee_\star x_2+(-1)^{|x_1||x_2|}b^\star_1(\bar R^\alpha(x_2))\vee_\star \bar R_\alpha(x_1)+b^\star_2(x_1,x_2))\nn\\
&=\sum_{i=0}^{3}b^\star_i(b^\star_0\vee_\star x_1\vee_\star x_2+b^\star_1(x_1)\vee_\star x_2+(-1)^{|x_1|}x_1\vee_\star b^\star_1(x_2)+b^\star_2(x_1,x_2))~.\nn\end{align}
In passing to the last line we assumed  the equivariance of the maps $b_i^\star$, i.e. we assumed  that they commute with the action of the twist generators. Thus one can show that the  homotopy relations are the same\footnote{Only when acting on explicit elements of the tensor algebra does one have to take into account the braided transposition map. In that case, the first difference with respect to the untwisted algebra appears when acting on three or more elements, as shown in \cite{Marija}.} as \eqref{bhom}. The braided coproduct \eqref{mapstar} and the compatible  coderivation \eqref{mapbimapstar} equivariant under the action of the degree zero twist element  reproduce,  in the coalgebra picture,   the braided $L_{\infty}$-algebra constructed in \cite{Marija}, c.p. Definition 4.73 in \cite{GR}.  Formally, the homotopy relations have the same form as \eqref{closed}:
\begin{equation}\label{closedstar}
\sum_{j=0}^i \sum_{\sigma\in \mathrm{Sh}(j,i)} b^\star_{i-j+1}(b^\star_j\vee_\star \mathrm{id}^{\vee_\star i})\circ\tau^\sigma_R=0~. \end{equation}
 Moreover, we have $\varepsilon\circ D=0$ as $\varepsilon(b_0(1))=0$.

The Hopf algebras $L_\infty^\star$ and $L_\infty^{\cal F}$ are isomorphic  and there exists an invertible map $\varphi$ between the underlying vector spaces \cite{thesis}
\bea\label{varphi}
{\varphi}(1)=1,\;{\varphi}(x)=\bar f^\alpha(x)\bar f_\alpha~,\eea
such that:
\begin{align}\label{mapD}
{\varphi}(x_1\vee_\star x_2)&={\varphi}(x_1)\vee {\varphi}(x_2)~,\\
 \Delta_\star&=({\varphi}^{-1}\otimes {\varphi}^{-1})\circ \Delta^{\cal F}\circ {\varphi}~,\label{mapDD}\\
 \varepsilon_\star&=\varepsilon\circ \varphi~,\label{mapco}\\
 S_\star&={\varphi}^{-1}\circ S\circ {\varphi}\label{mapS} ~.\end{align}

On the other hand, there exist maps between $L_{\infty}$-algebras: an $L_{\infty}$-morphism is a collection of graded symmetric maps $\phi=\{\phi_i:S^iX\to X',\;i\geq 0\}$ of degree zero from ${\bold S}(X)$ to ${\bold S}(X')$,  such that they define a coalgebra morphism i.e. satisfy:
\bea\label{copr1} 
\Delta'\circ \phi=(\phi\otimes \phi)\circ\Delta~,
\eea and such that $\phi$ is compatible with the coderivations:
\bea \label{copr2}
D'\circ\phi=\phi\circ D~.
\eea
The first few components are:
\begin{align}\label{mcomp}
	\phi(1)&=1+\phi_0+\sfrac 1{2!}\phi_0\vee' \phi_0+\cdots~,\nn\\
	\phi(x)&=\phi_1(x)+\phi_0\vee' \phi_1(x)+\sfrac 1{2!}\phi_0\vee'\phi_0\vee' \phi_1(x)+\cdots~,\nn\\
	\phi(x_1\vee x_2)&=\phi_1(x_1)\vee' \phi_1(x_2)+\phi_0\vee'\phi_1(x_1)\vee' \phi_1(x_2)+\cdots+\nn\\&\phantom{\,=\,}+\phi_2(x_1,x_2)+\phi_0\vee' \phi_2(x_1,x_2)+\cdots~.
\end{align}
When $\phi_0\neq 0$ we talk about curved $L_\infty$-morphisms.
From the compatibility of coderivations \eqref{copr2} one obtains the explicit relation between coderivation maps $b_i$ and $b_i'$ \cite{SK}
\begin{align}\label{morphc}
	\sum_{\s\in \mathrm{Sh}(l, n-l)}\!\!\!\phi_{1+l}\circ (b_{(n-l)}\otimes{\rm id}^{\otimes l})\circ\t^\s=\sum_{j=0}^\infty\sum_{k_1+\cdots+k_j=n}\sum_{\s\in\mathrm{Sh}(k_1,\ldots,k_j)}\mkern-20mu\sfrac1{j!}b'_j(\phi_{k_1}\vee\cdots\vee\phi_{k_j})\circ\t^\s~.
\end{align}
When the map $\phi_1$ is invertible, we have an $L_\infty$-isomorphism.
Applying the $L_\infty$-morphism to our case of interest, namely finding a map 
$\phi^\star: L_\infty^\star\to L_\infty^{\cal F}$, we need to define  the component maps $\phi_i^\star$   which are braided graded symmetric, i.e.
\bea
\phi_i^\star(x_1,\ldots,x_m,x_{m+1},\ldots,x_i)=(-1)^{|x_m||x_{m+1}|}\phi_i^\star(x_1,\ldots,\bar R^\a(x_{m+1}),\bar R_\a(x_{m}),\ldots,x_i)~,
\eea
and equivariant with respect to the action of twist generators. The expression for the morphism $\phi^\star$ is then obtained from \eqref{morphc} by exchanging $b_i$ with $b_i^\star$ and the  action of the permutation  $\t^\s$ with $\t^\s_R$. 

However,  things are much simpler  here; the Hopf algebra morphism is both an algebra morphism \eqref{mapD}  and a coalgebra morphism  (\ref{mapDD},~\ref{copr1}),  so we obtain  that the only non-vanishing component of the morphism $\phi^\star$ is  $\phi_1^\star$:
\bea
\phi_1^\star(x )=\varphi(x)=\bar f^\alpha(x)\bar f_\alpha~.
\eea
The  relation we established  between Hopf and  $L_{\infty}$-algebras implies that the morphism $\varphi$  between Hopf algebras \eqref{mapD}-\eqref{mapS} can be extended to a strict  $L_\infty$-morphism  by demanding compatibility of the morphism with the coderivation \eqref{copr2}
\bea
 \varphi\circ  D_{\star} &=&D_{\mathcal{F}} \circ \varphi~,\nn\\
 \varphi(b^\star_n(x_1,\ldots,x_n))&=&b_n(\varphi(x_1),\ldots,\varphi(x_n))~.
\eea
Notice that in Refs.\cite{cn1,cn2} the authors discussed the special example of twisting of an $L_\infty$-algebra, where the $L_\infty$-morphisms of twisted algebras went beyond strict $L_\infty$-morphisms. The difference comes from the difference between Hopf algebra modules which we discuss here and more general $L_{\infty}$-algebra modules, see \cite{LadaM,zambon}.

Finally,  in complete analogy with  Thm 2.1. we can relate   the braided $L_\infty$-algebra $(L_\infty^\star, D_\star)$ with the Hopf algebra $(L_\infty^\star, \vee_\star, \Delta_\star,\epsilon,S_\star)$. The compatibility relation between the antipode $S_\star$ and the codifferential $D_\star$ follows from the equivariance of the coderivation maps $b_i^\star$  and the fact that the antipode $S_\star$ is a graded algebra anti-homomorphism. 

\section{Concluding remarks}

In this paper we have identified the cocommutative and coassociative Hopf algebra structure underlying  $L_\infty$-algebras.  Thus we were able to introduce a non-(co)commutative deformation by applying the Drinfel'd twist approach \cite{dt} and obtaining  the braided $L_\infty$-algebra of Ref.\cite{Marija}  as a module of the twisted one.   In Ref.\cite{Marija} the braided $L_\infty$-algebra was used in the construction of a non-commutative deformation of the Chern-Simons and Einstein-Cartan-Palatini actions with a braided gauge symmetry.
However, the  physical interpretation    of braided gauge symmetries encountered in these models was not well understood. One  way to improve this situation is to construct an appropriate generalization of the BV formalism  \cite{AR,GR} that could help in identifying equivalent physical configurations. This is particularly natural in the coalgebra formulation, where one can interpret the dual of the codifferential as locally being a cohomological vector field $Q$ of degree 1 on a manifold $M$, i.e. $Q=D^*$ or:
$$Q=\sum_{i=0}^{\infty}\frac{1}{i!}C^\b_{\a_1...{\a_i}}z^{\a_1}\cdots z^{\a_i} \frac{\partial}{\partial z^\b}~. $$
Here, the   structure constants of the $L_{\infty}$-algebra are the components of the coderivation $D$ on a basis $\{\t_{\a} \}$  of  $X$:
$$ b_i(\t_{\a_1},...,\t_{\a_i})=C^\b_{\a_1...{\a_i}}\t_\b~, $$
and  $\{z^{\a}\}$  represent a basis of the dual\footnote{In the infinite-dimensional case one either restricts $X^\star$ to the space spanned by $\{z^{\a}\}$, or considers continuous duals in infinite-dimensional topological vector spaces, see discussion in  \cite{Alex}. } vector space $X^\star$.   In the BV formalism, $Q$ becomes the BRST operator and  $z^{\a}$ the physical fields.
 
Furthermore, in  the $L_\infty$-framework  there exists a well-defined notion, at least for flat $L_\infty$-algebras,  of an $L_\infty$-quasi-isomorphism that relates physically (gauge)  equivalent configurations. Namely, when the $0$-bracket vanishes, the $1$-bracket is a differential, see \eqref{bhom}, and there is a cochain complex underlying the $L_\infty$-algebra. In that case one defines  the $L_\infty$-quasi-isomorphisms by the requirement that the linear morphism component $\phi_1$   induces an isomorphism of  cohomologies of the respective $L_\infty$-algebras, see detailed discussion in \cite{Jurco}. For the case of a non-vanishing  $0$-bracket,  a natural setting would be that of $\s$-models and $L_\infty$-spaces introduced by Costello \cite{costello}. An $L_\infty$-space includes  target manifold data and  $0$-bracket can be identified with the curvature of a connection on the target. Said differently, the connection $x\in X$ is a degree zero solution of Maurer-Cartan equation 
$$\sum_{i=0}^\infty \sfrac{1}{i!}b_i(\underbrace{x,\ldots,x}_{i\;{\rm times}})=0~.$$
Using this solution one can define new $L_\infty$-algebra on the same vector space, but with vanishing curvature \cite{gg,cl}.

\paragraph{Acknowledgments.} We thank Marija Dimitrijevi\'c \'Ciri\'c and Peter Schupp for extensive discussions and Paolo Aschieri, Athanasios Chatzistavrakidis and Richard Szabo for useful comments. The work is supported by the Croatian Science Foundation project IP-2019-04-4168.

\appendix

\section{On $L_\infty$-algebras}
\begin{defn}{\rm ({\bf $L_\infty$-algebra \cite{ls}})}
 An $L_\infty$-algebra  $( {X},\mu_i)$ is a graded vector space $X$ equipped  with a collection of multilinear maps that are graded totally antisymmetric:
\be
\mu_i: {X}^{\otimes i}\to  {X}~, \nonumber
\ee
of degree $2-i$ where $i\in\mathbb{N}_0$ and  satisfy the homotopy Jacobi identities:
\be\sum_{j+k=n}\sum_\sigma\chi(\sigma;x)(-1)^{k}\mu_{k+1}(\mu_j(x_{\sigma(1)},\ldots,x_{\sigma(j)}),x_{\sigma(j+1)},\ldots,x_{\sigma(n)})=0~;\label{eq:homotopyjac}\nn\ee
for all $x_i\in {X}$,  $n\in\mathbb{N}_0$. Here $\chi(\sigma;l)$ indicates the graded Koszul sign including the sign from the parity of the permutation of $\{1,\ldots,n\}$ that is ordered as: $\sigma(1)<\cdots<\sigma(j)$ and $\sigma(j+1)<\cdots<\sigma(n)$.\label{linfty}
\end{defn}
  We use the convention that totally graded antisymmetric means:
\[\mu_i(\ldots,x_r,x_s,\ldots)=-(-1)^{|x_r||x_s|}\mu_i(\ldots,x_s,x_r,\ldots)~,\]
with $|x_r|$ the   degree of homogeneous element $x_r\in{ X}$.
When $\mu_0\neq 0$   this algebra is called a curved $L_\infty$-algebra, while the name flat $L_\infty$-algebra  refers to the case $\mu_0= 0$.

The homotopy Jacobi identities defining the $L_\infty$ structure exist for any given level $n$, and there can be, in principle, an infinite number of them. The first few homotopy relations are:
\begin{align} \label{lhom}
n&=0: && \mu_1\mu_0=0~,\nn\\
n&=1: && \mu_1^2(x)=\mu_2(\mu_0,x)~,\nn\\
n&=2: && \mu_1(\mu_2(x_1,x_2))-\mu_2(\mu_1(x_1),x_2)-(-1)^{1+|x_1||x_2|}\mu_2(\mu_1(x_2),x_1)=-\mu_3(\mu_0,x_1,x_2)~.\nn\end{align}
	These homotopy relations can be related to \eqref{bhom} using a degree $-1$ map $s$ between the algebra and coalgebra pictures called a suspension or shift isomorphism:  $$s:X\to X[1] \;\;{\rm  s.t.}\;\;  (X[1])_d=X_{d+1}~,$$
which  induces an isomorphism of the graded tensor algebras,
 \bea
 s^{\otimes i}:x_1\w\cdots\w x_i\to (-1)^{\sum_{j=1}^{i-1}(i-j)}sx_1\vee\cdots\vee sx_i~,\nn\eea
 and d\'ecalage isomorphism of the brackets:
 \bea
 \mu_i=(-1)^{\tfrac 12 i(i-1)+1}s^{-1}\circ b_i\circ s^{\otimes i}~.\nn
 \eea
\section{On  Hopf algebras}

\begin{defn}[Bialgebra]
	A bialgebra $(A,  \mu, \eta, \Delta, \varepsilon)$ over $K$ is a vector space  which is both an \textit{algebra} and a \textit{coalgebra} in a compatible way:
	\begin{equation}
		\Delta(hg)=\Delta(h)\Delta(g), \quad \Delta(1)=1\otimes1, \quad \varepsilon(hg)=\varepsilon(h)\varepsilon(g), \quad \varepsilon(1)=1,\quad  \forall h,g\in A.
	\end{equation}
	The comultiplication $\Delta:A\to A\otimes A$ and counit map $\varepsilon:A\to K$ are both algebra homomorphisms, whereas the multiplication $\mu:A\otimes A \to A$ and unit map $\eta: K\to A$ are coalgebra homomorphisms. 
\end{defn}
\begin{defn}[Hopf algebra]
	A Hopf algebra $(H,\mu,\Delta,\varepsilon, S)$ over $K$ is a bialgebra over $K$ equipped with an antipode map $S:H\to H$ satisfying the following:
	\begin{equation}
		\mu\circ (id\otimes S)\circ \Delta=\mu\circ (S\otimes id)\circ \Delta=\eta\circ \varepsilon~.
	\end{equation}
	
\end{defn}

\begin{rmk}
	If an antipode exists, it is unique \cite{Majid}.
\end{rmk} 
 The existence of an inverse antipode map $S^{-1}$ is not assumed, but if  $S^{2}=\text{id}$, the inverse is equivalent to the antipode map itself. 
 A consequence of the antipode's uniqueness is that it obeys the following relations $\forall h,g\in H$:
\begin{align}
	 S(hg)&=S(g)S(h)~,\mkern-500mu& S(1)&=1~,\\ 
	(S\otimes S)\circ\Delta (h) &= \Delta\circ S(h)~, \mkern-100mu&\varepsilon S(h)&=\varepsilon (h)~. 
\end{align}
The first two relations state that the antipode is an antialgebra map, whereas the second two state that it is an anticoalgebra map.

\end{document}